Role of proton irradiation and relative air humidity on iron corrosion

S. Lapuerta[1,2], N. Moncoffre[1], N. Millard-Pinard[1], H. Jaffrézic[1], N. Bérerd[1], D. Crusset[2]

[1] Institut de Physique Nucléaire de Lyon , 4, rue Enrico Fermi , 69622 Villeurbanne cedex,
[2] ANDRA Parc de la Croix Blanche 1-7 rue Jean Monnet, F-92298 Châtenay-Malabry Cedex, France

Abstract

This paper presents a study of the effects of proton irradiation on iron corrosion. Since it is known that in humid atmospheres, iron corrosion is enhanced by the double influence of air and humidity, we studied the iron corrosion under irradiation with a 45% relative humidity. Three proton beam intensities (5, 10 and 20 nA) were used. To characterise the corrosion layer, we used ion beam methods (Rutherford Backscattering Spectrometry (RBS), Elastic Recoil Detection Analysis (ERDA)) and X-ray Diffraction (XRD) analysis. The corrosion kinetics are plotted for each proton flux. A diffusion model of the oxidant species is proposed, taking into account the fact that the flux through the surface is dependent on the kinetic factor *K*. This model provides evidence for the dependence of the diffusion coefficient, *D*, and the kinetic factor, *K*, on the proton beam intensity. Comparison of the values for *D* with the diffusion coefficients for thermal oxygen diffusion in iron at 300 K suggests an enhancement due to irradiation of 6 orders of magnitude.

**Corresponding Author :** Sandrine Lapuerta

**Full Mailing Address :**
IPNL
4 rue Enrico Fermi
Bat. Paul Dirac
69622 Villeurbanne Cedex

**Telephone :** +33 4 72 43 10 63
**Fax** : +33 4 72 44 80 04
**E-mail :** lapuerta@ipnl.in2p3.fr

## 1. Introduction

In the context of long term geological storage, high level nuclear wastes will be placed in stainless steel containers surrounded by a low alloyed carbon steel overpack as a second barrier. These containers will be exposed to humid air and mainly to $\gamma$ irradiation.

In this context, we propose a fundamental study to better understand the corrosion mechanisms of pure iron, considered as a model material, under proton irradiation. The irradiation effect was studied by using a 3 MeV proton beam characterised mainly by an electronic stopping power. We have chosen proton irradiation for two main reasons. First, the protons produce a high ionisation density correlated with a high electronic stopping power compared to $\gamma$ irradiation. Second, the beam energy control allows to study the corrosion process precisely at the iron/atmosphere interface.

In a recent review paper [1], the atmospheric corrosion mechanisms occurring inside a rust layer during a wet-dry cycle were presented. This modelling demonstrated the enhancement of iron corrosion by the joint influence of oxygen and water.

The aim of the present work is to study the iron corrosion kinetics in air with a relative humidity (*RH*) of 45% and under proton irradiation. We will first present evidence for oxygen and hydrogen migration in the corrosion layer which was characterised by Rutherford Backscattering Spectrometry (RBS) and Elastic Recoil Detection Analysis (ERDA). Then, in order to attempt to identify the crystalline phases in the corrosion products formed on the iron samples, the results of X-Ray Diffraction (XRD) characterisation will be presented. Finally, we will focus on the modelling of the corrosion iron kinetics as a function of the proton beam intensities.

## 2. Experimental

### 2.1 Irradiation conditions

The studied material was a thin iron foil (99.85% purity), 10 μm thick, provided by Goodfellow. This foil could not be polished because it was too thin. In the following, it will be referred to as the initial sample. The irradiation experiments were performed using the 4 MV Van de Graaff accelerator of the Nuclear Physics Institute of Lyon (IPNL) which delivers a 3 MeV proton beam characterised mainly by an electronic energy deposition (the linear energy transfer is equal to $3x10^{-2}$ keV $\mu m^{-1}$). The proton beam is extracted from the beam line vacuum ($10^{-3}$ Pa) to the atmosphere by crossing a 5 μm thick havar (Co/Cr/Ni alloy) window. The external proton beam enters the irradiation cell through the studied iron foil and stops in water. The experimental set up is displayed in Fig. 1. The relative humidity of the air in the 8 mm thick gap between the havar window and the iron foil was controlled and measured throughout the experiment using a Hygropalm humidity controller. The measured *RH* and temperature were 45 ± 2 % and 300 K respectively. The irradiation line was equipped with a sweeping system which allowed a homogeneous irradiation over the whole iron surface (5x5 mm). During irradiation, the beam intensity was constant and measured carefully with a beam chopper that had previously been calibrated. Three intensities (5, 10 and 20 nA) were used.

### 2.2 Analysis conditions

For each irradiation condition, the iron surface in contact with the humid air was analysed by using ion beam analysis performed on the IPNL 4 MV Van de Graaff accelerator. RBS coupled with ERDA were used to determine the corrosion layer evolution. For RBS analysis, a 172° detection angle was used while the alpha incident particles energy was adjusted to 1.7 or 3 MeV depending on the oxide layer thickness. ERDA analysis was induced using 1.7 MeV α particles. The incident angle on the target and the detection angle were respectively 15° and 30°. A 6.5 µm thick polymer (mylar) absorber was placed in front of the Si detector to stop the backscattered α particles. In such conditions, the depth resolution was close to 20 nm in iron.

The SIMNRA program was used to simulate the energy spectra obtained both by RBS and ERDA so as to determine the atomic concentration profiles of iron, oxygen and hydrogen.

## 3. Characterisation of the corrosion layers induced by irradiation

Oxygen and hydrogen profiles are presented respectively in Fig. 2 and 3 both for the initial sample and the irradiated ones. The oxygen profiles show a continuous decrease with increasing depth, both for the initial sample and the irradiated ones. The irradiation effect is clearly shown by the strong oxygen concentration enhancement. In contrast, the hydrogen profile corresponding to the initial sample displays a very superficial contamination whereas the irradiated ones show an uniform and high concentration level extending into the iron.

In order to discuss the stoichiometry variation of this corrosion layer, we have plotted in Fig.4 the $C_O/C_{Fe}$ atomic concentration ratio as a function of the irradiation time at three different depths: the surface (Fig. 4a), at 50 nm (Fig. 4b) and at 100 nm (Fig. 4c). Each ratio was calculated for a step equal to the resolution depth. The $C_O/C_{Fe}$ ratio value corresponding to the chosen depth in the initial sample is shown by a black spot located at $t = 0$. For each depth, these ratios are compared to the FeOOH and FeO stoichiometries which are the two extreme iron oxidation states of rust (i.e. Fe(II) and Fe(III)). At the surface (Fig. 4a), for the 10 and 20 nA beam intensities, we observe that after an irradiation of 30 minutes the $C_O/C_{Fe}$ ratio stands between 1.5 and 2; whereas at 50 nm it stands around 1 (Fig. 4b) and at 100 nm it is below 1 (Fig. 4c). In the case of the 5 nA irradiation intensity, the $C_O/C_{Fe}$ ratio was systematically lower than 1 but higher than that of the initial sample. In addition, we have plotted in figure 5 the $C_H/C_O$ atomic concentration ratio as a function of the irradiation time at three different depths: the surface (Fig. 5a), a 50 nm depth (Fig. 5b) and a 100 nm depth (Fig. 5c). These ratios are compared to the FeOOH stoichiometry. The $C_H/C_O$ ratio corresponding to the initial sample is shown by a black spot at $t$ = 0. At the surface, it quickly reaches a value of 0.4, whatever the irradiation intensity. At the depths of 50 nm and 100 nm, the ratio value is about 0.25 irrespective the irradiation. It shows that the $C_H/C_O$ ratio is always lower than the one of FeOOH. From these results, we can assume that the corrosion product is composed of mixed hydroxides and oxides as previously observed [3].

Classical X-ray diffraction experiments were performed on the 45 min irradiated samples and on the initial sample and the results are presented in Fig. 6. The main peaks of the $Fe_3O_4$ and β-FeOOH phases [2] are marked by arrows on this figure. On the initial sample (Fig. 6a), we observe a large diffraction peak centred at a $2\theta$ angle of 18° which corresponds to the surface oxide layer. From the large width of the X-ray diffraction peaks, we can assume that the layer structure is almost amorphous. On Fig. 6b, the X-ray diffraction pattern corresponding to the samples irradiated at different intensities are presented. From this comparison, it appears that as the incident proton flux increases, a balance occurs between two crystallographic structures which correspond to the peaks observed respectively at 18° and 36°. At a beam intensity of 5 nA, only the 18° contribution is observed. At 10 and 20 nA, this peak progressively vanishes and the peak at 36° increases.

However, these results can only give some hints to the mineral phases but are insufficient to clearly identify these phases.

## 4. Modelling the corrosion kinetics

In this part, we focus on the kinetics of oxide growth during the corrosion process induced by irradiation. In order to identify the oxygen migration mechanism, we have used the relation:

$m_O(t) = A\, t^n$ (1)

where $m_O(t)$ is the oxygen mass gain, $t$ the irradiation time and $A$ and $n$ are constants [3].

As shown in Fig. 7, which displays log $m_O(t)$ versus log $t$, with the experimental results represented by dots, the straight lines obtained have the same slopes, whatever the irradiation intensity. Hence, the corrosion mechanism does not depend on the beam intensity. The value of n deduced from Fig. 7 is 0.75 ± 0.07 with $R^2 = 0.98$. Therefore we applied a model based on the Crank mathematical resolution of the Fick's equations [4,5] which has recently been used in the case of zirconium oxidation under irradiation [6]. Fick's second law equation was applied by considering that the diffusion coefficient $D$ is independent of the oxygen concentration $C(x,t)$.

$$\frac{\partial C(x,t)}{dt} = D\frac{\partial^2 C(x,t)}{\partial x^2} \qquad (2)$$

The initial condition takes into account the fact that the oxygen concentration in the iron bulk at $t = 0$ is $C(x,0) = 0$ (3)

As a result on the irradiation process, the oxygen atomic concentration at the surface is time dependent. The flux of oxygen entry into the iron surface is expressed as:

$J(0,t) = K[C_s^0 - C_s(t)]$, (4)

where $K$ is the kinetic constant of the oxygen surface exchange, $C_s(t)$ is the oxygen concentration at the surface at a time $t$, and $C_s^0$ is the oxygen concentration at equilibrium. Our experimental results show that the surface composition is a mixture of oxide and hydroxide phases which composition depends on the irradiation conditions. Indeed, if we consider two extreme cases of composition ratio (100% for $Fe_3O_4$ or 100% for FeOOH), the corresponding $C_s^0$ values remain very close and respectively equal to 5.2x10$^{22}$ cm$^{-3}$ and 5.8x10$^{22}$ at cm$^{-3}$. Hence, a mean value is taken in the calculation of the diffusion coefficient corresponding to a mixture of 50 at.% for $Fe_3O_4$ and 50 at.% for FeOOH. This approach allows calculation of the oxygen concentration as function of depth and time and expression of *J(0,t)*. The rate of increase of the mass of oxygen in the metal due to irradiation, $m_O(t)$, is given by:

$$m_O(t) = \int_0^t J(0,t)dt = \frac{c_s^0}{h}\left[\exp(h^2Dt)erfc(h\sqrt{Dt} - 1 + \frac{2}{\sqrt{\pi}}h\sqrt{Dt}\right] \text{ [4] where } h=K/D \qquad (5)$$

A comparison between the theoretical and experimental values is given in Fig. 8 where M(t) is plotted versus time. The *D* and *K* values deduced from this fit are given in Table 1. The accuracy is estimated to be 20%. Comparison of the values for *D* in Table 1 with the diffusion coefficients for thermal oxygen diffusion in iron at 300 K [7] suggests an enhancement due to irradiation of 6 orders of magnitude

For the irradiation intensities of 10 and 20 nA, the *K* values were the same and correlated to a $C_H/C_O$ surface ratio close to 0.4. In case of the 5 nA irradiation intensity, the *K* value was smaller as expected by the large $C_H/C_O$ surface value (0.9).

The oxygen diffusion coefficients increase classically with the beam intensity [5]. Indeed the diffusion process is controlled by the density of defects produced by irradiation.

## 5. Discussion

Under irradiation, $O_2^\bullet$ and $HO^\bullet$ radicals amongst others are created in humid air. Their radiolytic yield is given in the literature [8,9]. Under our irradiation conditions, the calculated concentrations of these radicals do not allow an explanation of the important oxidation process observed. In order to explain it, we can assume that there are adsorbed species at the iron surface and that the proton beam enhances significantly the migration of these species as well as their interactions. The excess of hydrogen can be explained by the formation of $H^+(H_2O)_n$ clusters, as already discussed in the literature [10], which shows that 80% of the charged species are in the form $H^+(H_2O)_n$ as soon as 1 at.% of vapour water is present in the air. This relative humidity corresponds to our experimental conditions.

Table I: *D* and *K* values calculated by the theoretical approach for the three beam intensities.

| Intensity (nA) | 5 | 10 | 20 |
|---|---|---|---|
| $D$ ($cm^2.s^{-1}$) | $2x10^{-13}$ | $4x10^{-13}$ | $4x10^{-12}$ |
| $K$ ($cm.s^{-1}$) | $9x10^{-9}$ | $2x10^{-8}$ | $2x10^{-8}$ |

Figure captions

Figure 1: Schematic representation of the set up for the irradiation experiments.

Figure 2 : Oxygen profiles deduced from RBS measurements for the initial sample and for the irradiated samples after 45 minute exposures at beam intensities of 5, 10 and 20 nA. The nm depth scale is calculated assuming pure iron ($\rho$ = 7.86 g.cm$^{-3}$). The error bars are represented on the first 150 nm depth on which both oxygen and hydrogen have been analysed.

Figure 3: Hydrogen profiles deduced from ERDA measurements for the initial sample and for the irradiated samples after 45 minute exposures at beam intensities of 5, 10 and 20 nA. The nm depth scale is calculated assuming pure iron ($\rho$ = 7.86 g.cm$^{-3}$).

Figure 4: Representation of the $C_O/C_{Fe}$ atomic concentration ratio as a function of the irradiation time for the irradiated samples at the surface (a), at 50 nm depth (b) and at 100 nm depth (c). The $C_O/C_{Fe}$ ratio value corresponding to the initial sample is shown by a black spot at $t = 0$. The dots sizes are representative of the error bars (2 to 5 %).

Figure 5: Representation of the $C_H/C_O$ ratio as a function of the irradiation time for the irradiated samples at the surface (a), at 50 nm depth (b) and at 100 nm depth (c). The $C_H/C_O$ ratio value corresponding to the initial sample is shown by a black spot at $t = 0$. The dots sizes are representative of the error bars (2 to 5 %).

Figure 6: X ray diffraction spectra for the initial sample (a) and for irradiated samples (b). The Cu K$\alpha$ radiation is used. X-rays diffraction peak positions corresponding to the $Fe_3O_4$ and $\beta$-FeOOH phases are indicated by arrows.

Figure 7: Representation of log $m_O(t)$ as a function of log($t$) where $m_O$ is in cm$^{-2}$ and $t$ in min. Experimental results are shown by dots. The full lines suppose a linear fit regression.

Figure 8: Oxygen gain as function of irradiation time. Experimental results are represented by dots and the full lines correspond to the theoretical fit.

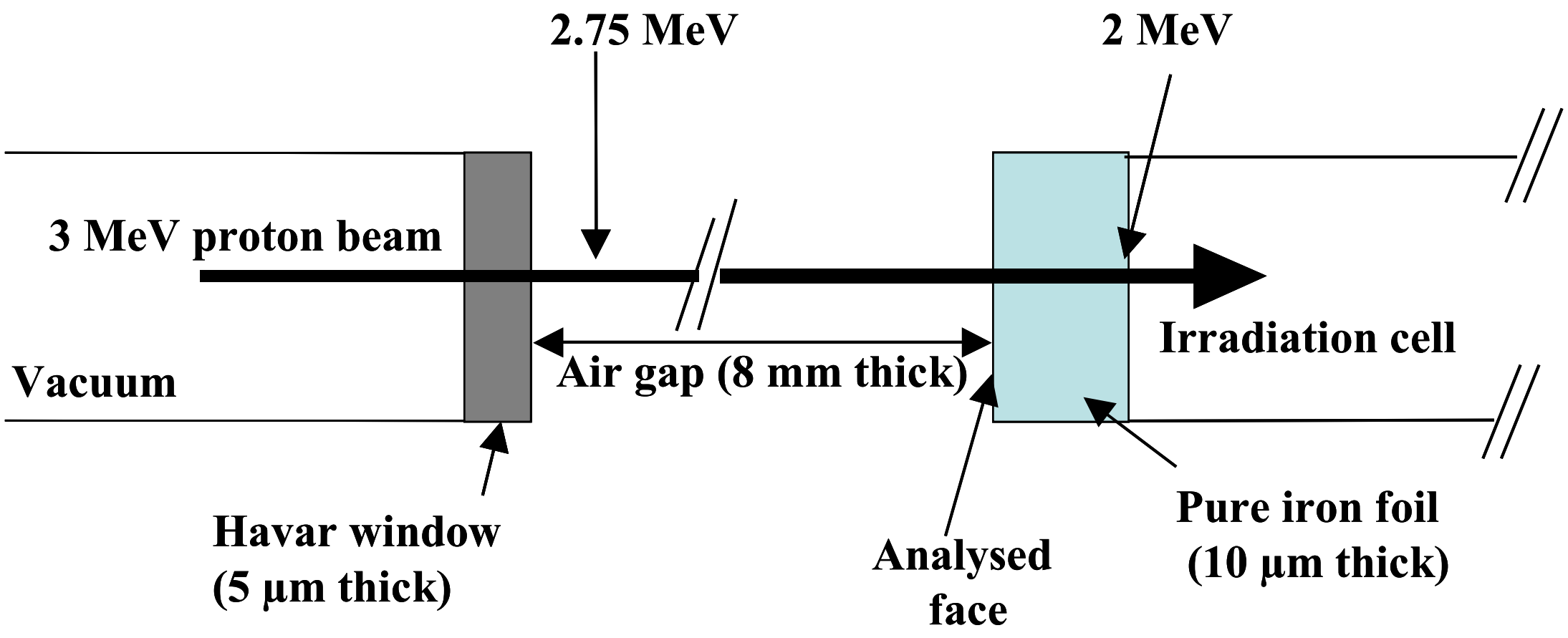

Figure 1

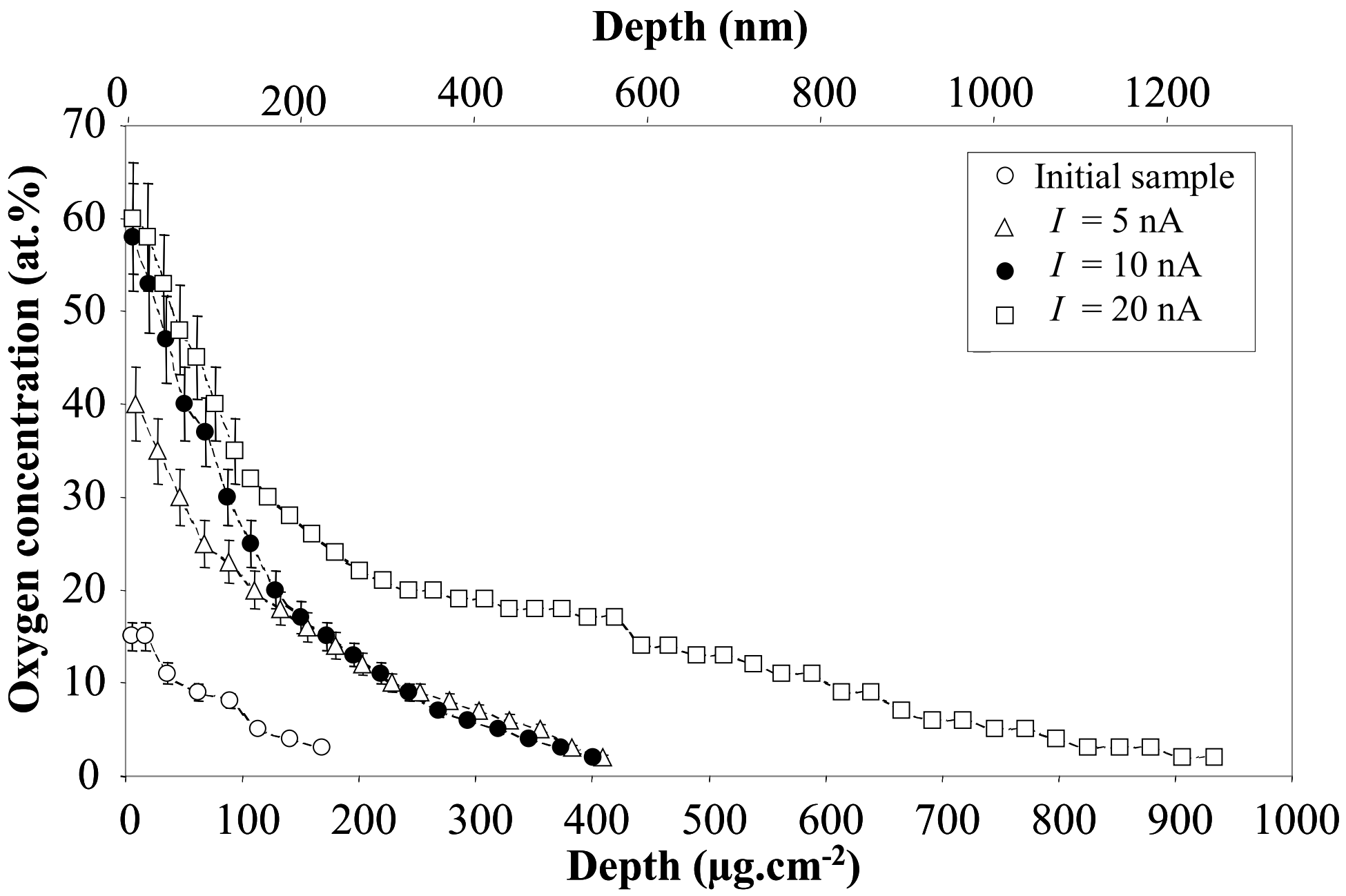

Figure 2

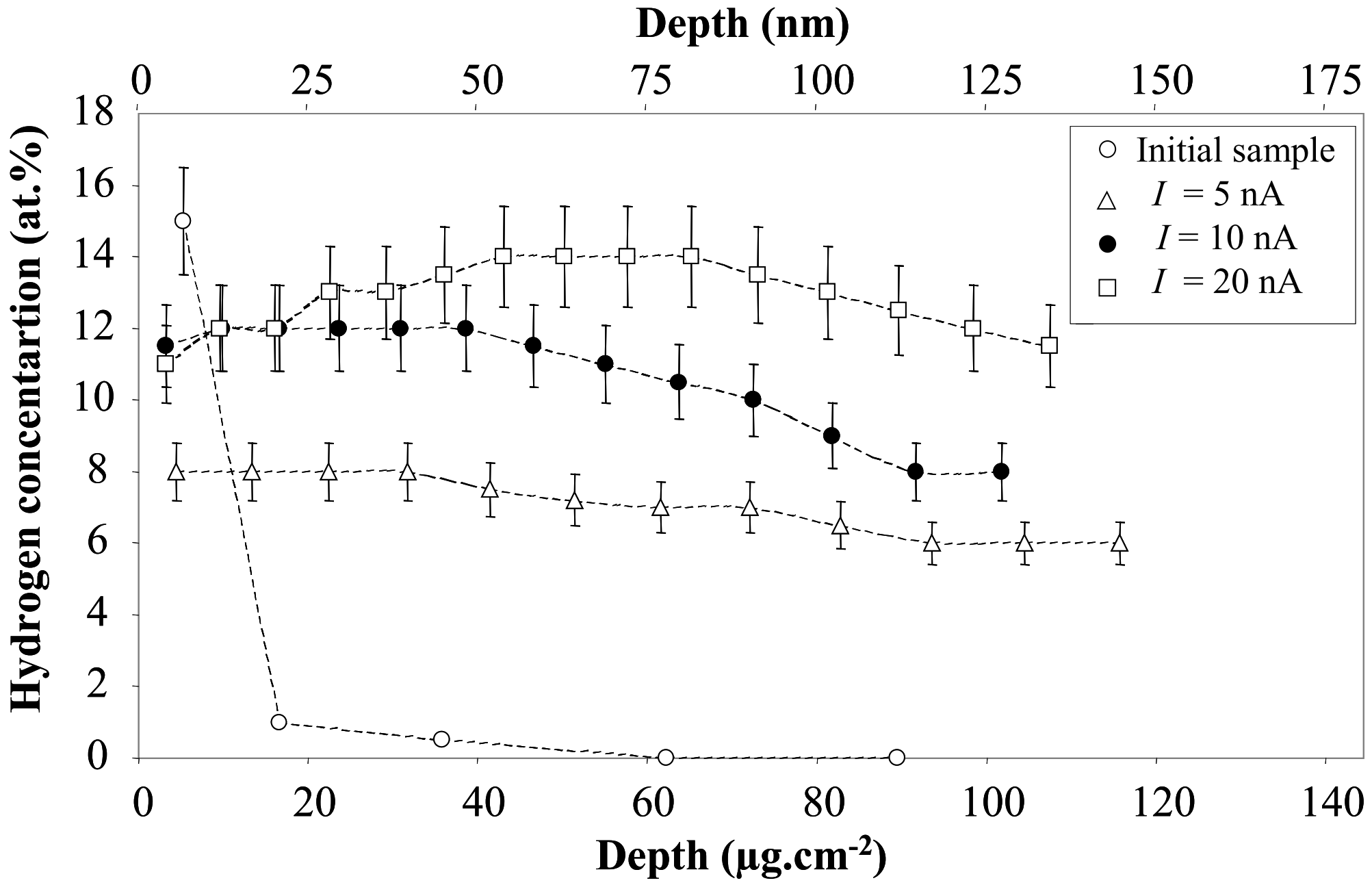

Figure 3

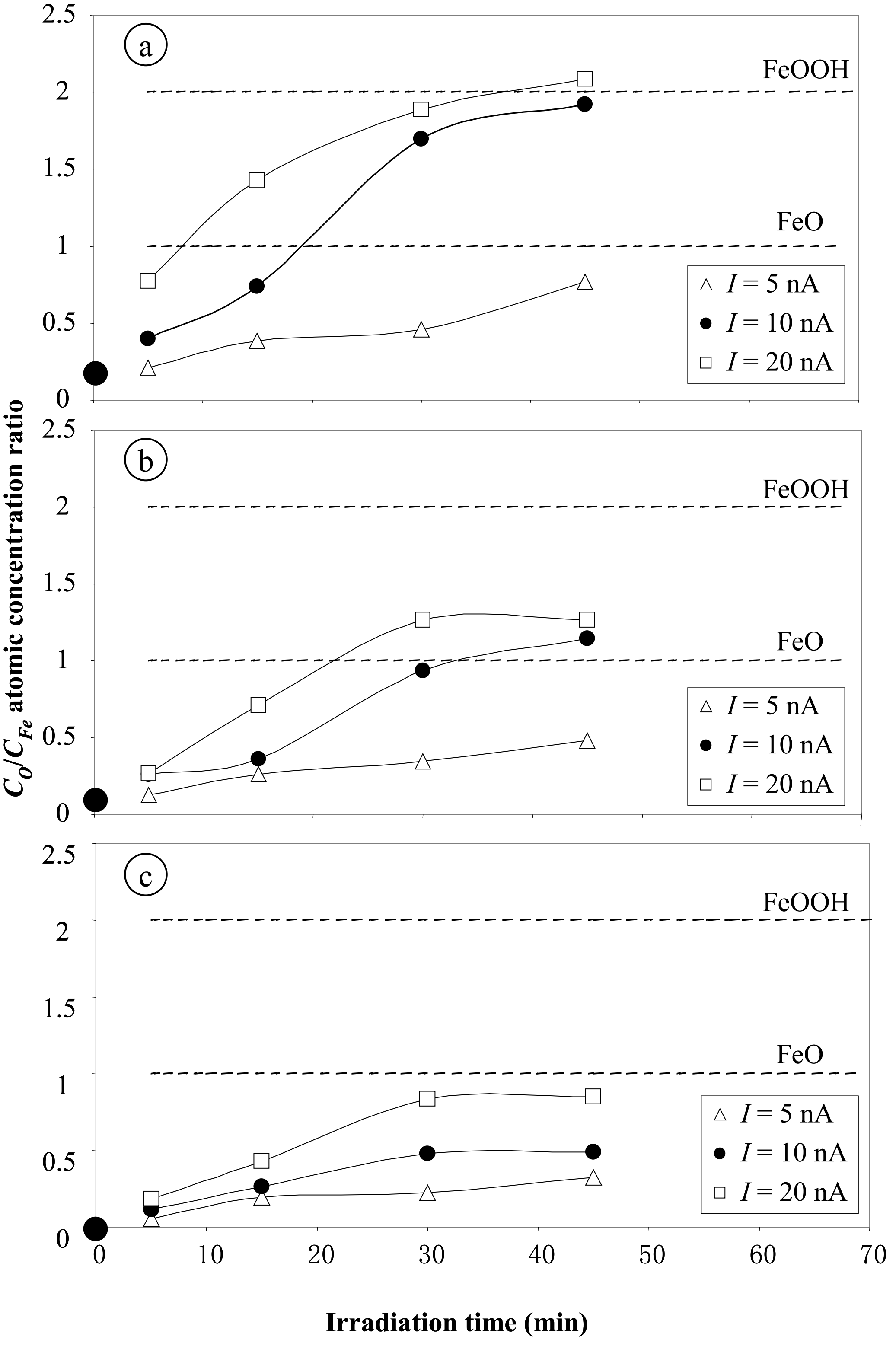

Figure 4

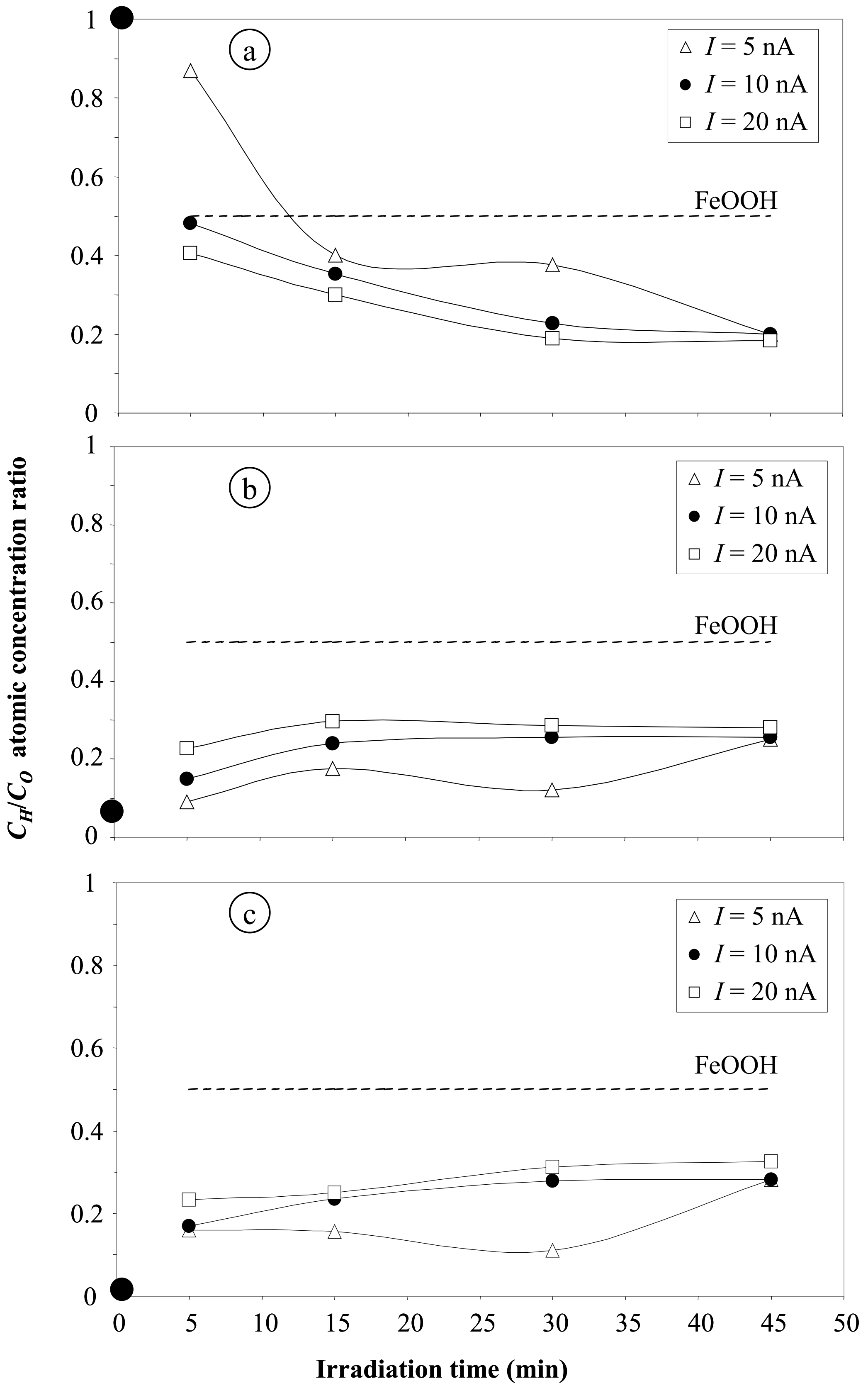

Figure 5

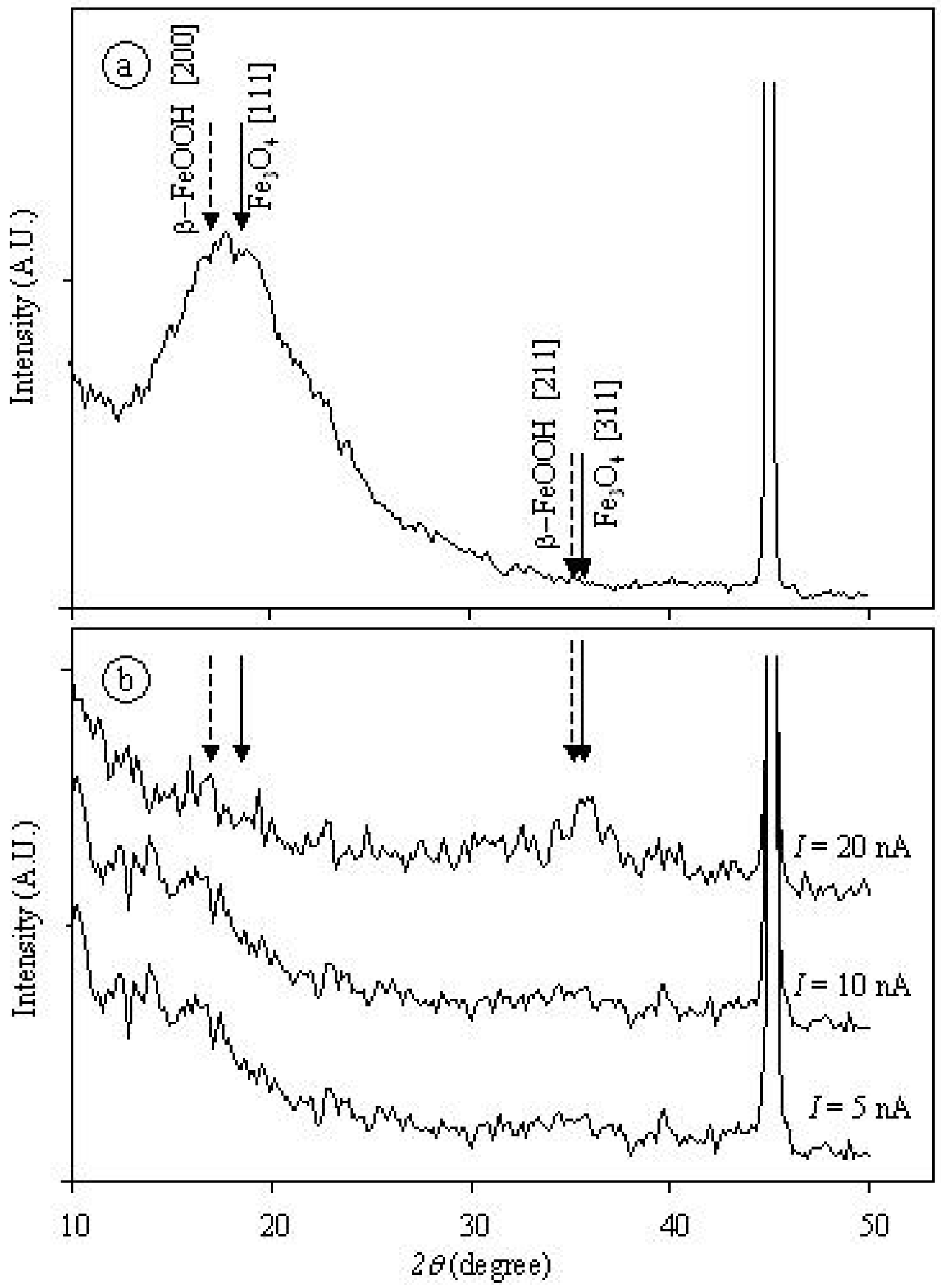

Figure 6

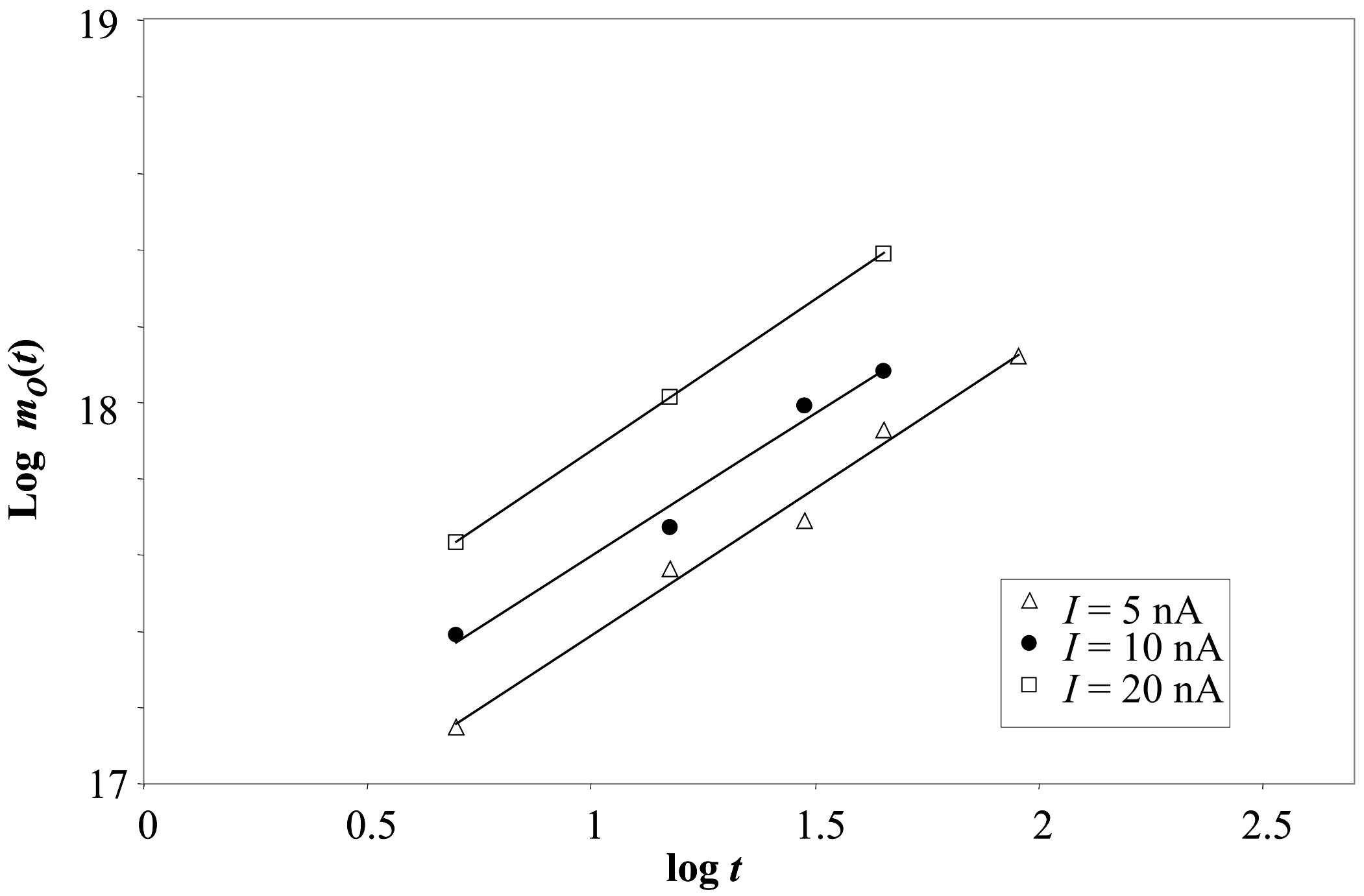

Figure 7

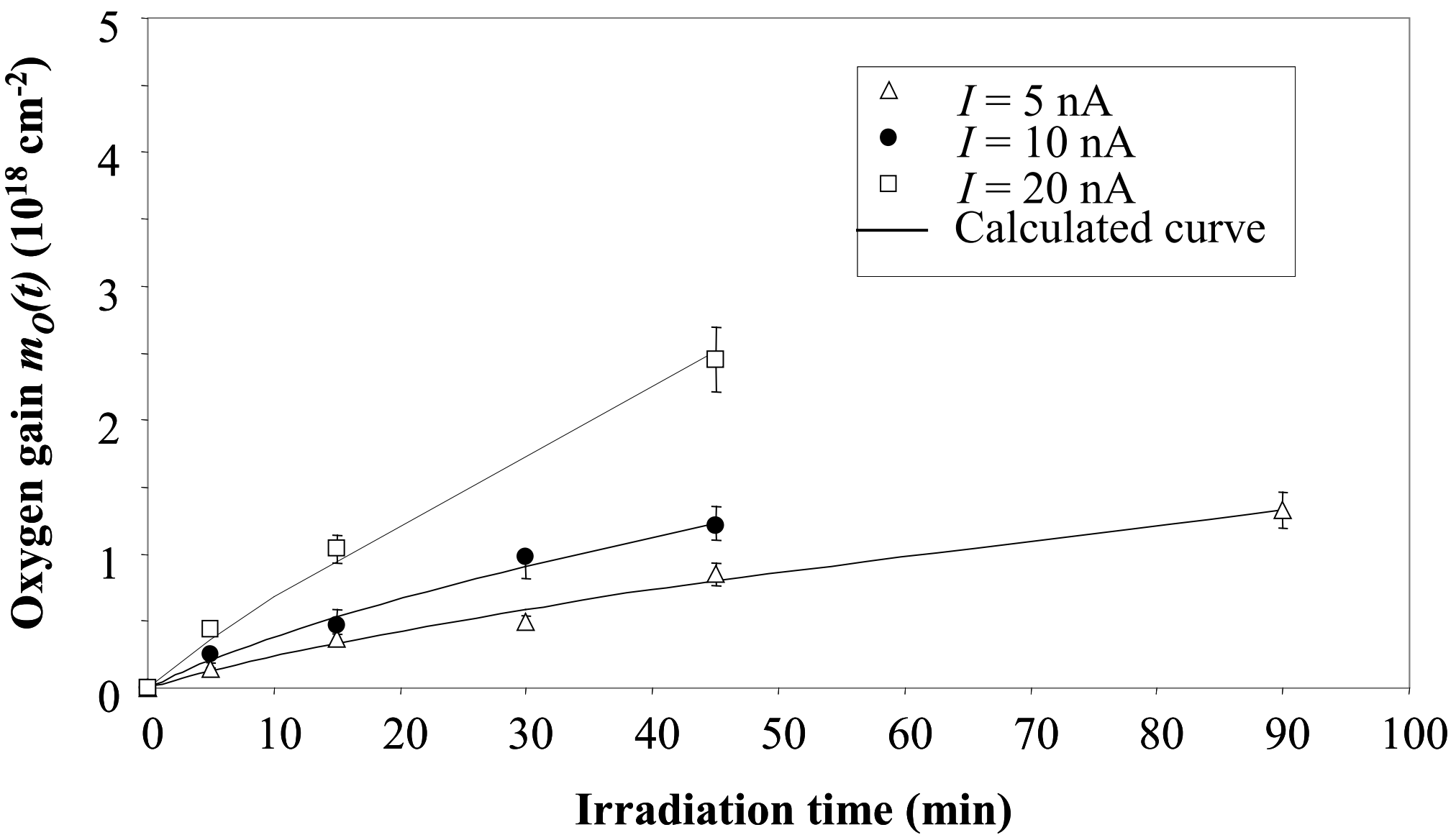

Figure 8